\newcommand{\beq}{\begin{equation}}
\newcommand{\eeq}{\end{equation}}
\newcommand{\bea}{\begin{eqnarray}}
\newcommand{\eea}{\end{eqnarray}}

\newcommand{\gsim}{\lower.7ex\hbox{$\;\stackrel{\textstyle>}{\sim}\;$}}
\newcommand{\lsim}{\lower.7ex\hbox{$\;\stackrel{\textstyle<}{\sim}\;$}}

\documentclass[aps,prev,twocolumn,preprintnumbers,floatfix,nofootinbib]{revtex4-1}
\usepackage{graphicx}
\usepackage{epstopdf}
\usepackage{mathrsfs}
\usepackage{amssymb}
\usepackage{verbatim}
\usepackage{color}
\usepackage{multirow}

\def\stacksymbols #1#2#3#4{\def\theguybelow{#2}
    \def\vp{\lower#3pt}
    \def\sp{\baselineskip0pt\lineskip#4pt}
    \mathrel{\mathpalette\intermediary#1}}

\def\intermediary#1#2{\vp\vbox{\sp
     \everycr={}\tabskip0pt
     \halign{$\mathsurround0pt#1\hfil##\hfil$\crcr#2\crcr
              \theguybelow\crcr}}}

\def\be{\begin{equation}}
\def\ee{\end{equation}}
\def\bea{\begin{eqnarray}}
\def\eea{\end{eqnarray}}

\def\sp{\;\;\;,\;\;\;}

\def\lsim{\raise0.3ex\hbox{$\;<$\kern-0.75em\raise-1.1ex\hbox{$\sim\;$}}}
\def\gsim{\raise0.3ex\hbox{$\;>$\kern-0.75em\raise-1.1ex\hbox{$\sim\;$}}}

\def\inbar{\,\vrule height1.5ex width.4pt depth0pt}

\def\IC{\relax\hbox{$\inbar\kern-.3em{\rm C}$}}
\def\IQ{\relax\hbox{$\inbar\kern-.3em{\rm Q}$}}
\def\IR{\relax{\rm I\kern-.18em R}}
 \font\cmss=cmss10 \font\cmsss=cmss10 at 7pt
\def\IZ{\relax\ifmmode\mathchoice
 {\hbox{\cmss Z\kern-.4em Z}}{\hbox{\cmss Z\kern-.4em Z}}
 {\lower.9pt\hbox{\cmsss Z\kern-.4em Z}}
 {\lower1.2pt\hbox{\cmsss Z\kern-.4em Z}}\else{\cmss Z\kern-.4em Z}\fi}

\def\comment#1{}

\def\u1x{U(1)_X}
\newcommand{\nc}{\newcommand}
\nc{\LL}{L}
\nc{\vv}{\tilde{v}}
\nc{\ccdot}{\!\cdot\!}
\nc{\gsm}{G_{SM}}
\nc{\vfive}{\mathbf{5}\oplus\mathbf{\overline{5}}}
\nc{\vten}{\mathbf{10}\oplus\mathbf{\overline{10}}}
\nc{\zhol}{Z^{\rm hol}}
\nc{\xfb}{\,{\rm fb}}

\begin{document}

%
%

\preprint{KCL-PH-TH/2013-15,LCTS/2013-09,CERN-PH-TH/2013-089,ACT-3-13,MIFPA-13-16,UMN--TH--3204/13,FTPI--MINN--13/15}

\vspace*{1mm}

\title{No-Scale Supergravity Realization of the Starobinsky Model of Inflation}

\author{John~Ellis$^{a}$}
\email{John.Ellis@cern.ch}
\author{Dimitri~V.~Nanopoulos$^{b}$}
\email{dimitri@physics.tamu.edu}
\author{Keith A. Olive$^{c}$}
\email{olive@physics.umn.edu}

\vspace{0.1cm}
\affiliation{
${}^a$ Theoretical Particle Physics and Cosmology Group, Department of
  Physics, King's~College~London, London WC2R 2LS, United Kingdom;\\
Theory Division, CERN, CH-1211 Geneva 23,
  Switzerland
 }
 \affiliation{
${}^b$ 
 George P. and Cynthia W. Mitchell Institute for Fundamental Physics and Astronomy,
Texas A\&M University, College Station, TX 77843, USA;\\
Astroparticle Physics Group, Houston Advanced Research Center (HARC), Mitchell Campus, Woodlands, TX 77381, USA;\\
Academy of Athens, Division of Natural Sciences,
28 Panepistimiou Avenue, Athens 10679, Greece,}
 
\affiliation{
${}^c$ 
William I. Fine Theoretical Physics Institute, School of Physics and Astronomy,\\
University of Minnesota, Minneapolis, MN 55455,\,USA}

\begin{abstract} 

We present a model for cosmological inflation based on a no-scale supergravity sector
with an SU(2,1)/SU(2) $\times$ U(1) K\"ahler potential, a single modulus $T$ and an inflaton superfield $\Phi$
described by a Wess-Zumino model with superpotential parameters $(\mu, \lambda)$. When $T$ is fixed,
this model yields a scalar spectral index
$n_s$ and a tensor-to-scalar ratio $r$ that are compatible with the Planck
measurements for values of $\lambda \simeq \mu/3M_P$. For the specific choice $\lambda = \mu/3M_P$, 
the model is a no-scale supergravity realization of the $R+R^2$ Starobinsky model.
\end{abstract}

\maketitle


\setcounter{equation}{0}




The initial release of cosmic microwave background (CMB) data from the Planck
satellite~\cite{Planckinflation} confront theorists of cosmological inflation~\cite{oliverev,inflationreviews} with a challenge. On the one
hand, the data have many important features that are predicted qualitatively by
the inflationary paradigm. For example, there are no significant signs of
non-Gaussian fluctuations or hints of non-trivial topological features such as
cosmic strings, and the spectrum of scalar density perturbations exhibits a
significant tilt: $n_s \simeq 0.960 \pm 0.007$, as would be expected if the effective scalar energy density
decreased gradually during inflation. On the other hand, many previously
popular field-theoretical models of inflation are ruled out by a combination of the
constraint on $n_s$ and the tensor-to-scalar ratio $r < 0.08$ as now imposed by
Planck {\it et al}: see, e.g., Fig.~1 of~\cite{Planckinflation}.
The only model with truly successful predictions
displayed in Fig.~1 of~\cite{Planckinflation} is the $R^2$ inflation model of Starobinsky~\cite{R2},
though similar predictions are made in Higgs inflation~\cite{HI} and related models~\cite{others}.

In the following paragraphs we provide the approach to inflation taken in this paper, which
casts a new light on the Starobinsky model~\cite{R2} and embeds it in a more general theoretical
context that connects with other ideas in particle physics. Specifically, the upper
limit on $r$ implies that the energy scale during inflation must be much smaller than
the Planck energy $\sim 10^{19}$~GeV. Such a hierarchy of energy scales can be
maintained naturally, without fine-tuning, in a theory with supersymmetry~\cite{ENOT}.
As is well known, (approximate) supersymmetry has many attractive features, such as
providing a natural candidate for dark matter and facilitating grand unification,
as well as alleviating the fine-tuning of the electroweak scale. In the context of
early-Universe cosmology, one must combine supersymmetry with gravity via a suitable 
supergravity theory~\cite{sugra}, which should accommodate an effective inflationary
potential that varies slowly over a large range of inflaton field values.
This occurs naturally in a particular class of supergravity models~\cite{no-scale}, 
which are called `no-scale'
because the scale at which supersymmetry is broken is undetermined in a first approximation,
and the energy scale of the effective potential can be naturally much smaller than $\sim 10^{19}$~GeV,
as required by the CMB data. No-scale models have the additional attractive feature that
they arise in generic four-dimensional reductions of string theory~\cite{Witten}, though this does not play
an essential r\^ole in our analysis. The attractive features of this no-scale supergravity
framework for inflation do not depend sensitively on the supersymmetry-breaking scale, which
could be anywhere between the experimental lower limit $\sim 1$~TeV from the LHC~\cite{MC} and 
$\sim 10^{10}$~TeV from the tensor-to-scalar ratio.

We now discuss these motivations at greater length before entering into the details of
our inflationary model.

Since the energy scale during the
inflationary epoch is typically $\ll M_P$, it is natural to study renormalizable
models, i.e., some combination of $\phi^2$, $\phi^3$ and $\phi^4$ in the single-field case.
In this spirit, it was shown in~\cite{KLtest,Croon} that a single-field model with a
potential of the form
\begin{equation}
V \; = \; A \phi^2 (v - \phi)^2
\label{model}
\end{equation}
could easily produce Planck-compatible values of $(n_s, r)$ for a suitable number of e-folds
before the end of inflation $N \sim 50$ to 60.
This simple symmetry breaking potential has a long pedigree, having been proposed initially
in~\cite{ab} (for a review, see~\cite{oliverev}) where it was argued that successful inflation
would require a small value of $A$ and $v > M_P$.

As we pointed out in~\cite{ENOT}, in addition to all the well-known reasons for postulating
low-scale supersymmetry, the small values of the quartic and quadratic
couplings that would be required in a successful inflationary model, e.g., $A$ in the above example,
become technically natural in the presence of low-scale supersymmetry.
In particular, small values of $\delta \rho/\rho$ become technically natural 
if approximate supersymmetry is invoked~\cite{ENOT}, and if the GUT Higgs is distinguished from the
singlet field that produces inflation, that later became known as the inflaton~\cite{nos}.

The simplest globally-supersymmetric model is the Wess-Zumino model
with a single chiral superfield $\Phi$~\cite{WZ}, which is characterized by a mass
term $ \hat \mu$ and a trilinear coupling $\lambda$, with the superpotential
\begin{equation}
W \; = \; \frac{\hat \mu}{2} \Phi^2 - \frac{\lambda}{3} \Phi^3 \, .
\label{WZW}
\end{equation}
As was discussed in~\cite{Croon}, the effective potential of the Wess-Zumino model 
reduces to (\ref{model}) when the imaginary part of the scalar component of $\Phi$
vanishes, in which case this model yields Planck-compatible inflation for a suitable
small value of $\lambda$.

However, global symmetry is not enough. As discussed above, in the context of early-Universe
cosmology one should certainly include gravity and hence construct a
locally-supersymmetric model, i.e., upgrade to supergravity~\cite{sugra}. 
The first attempt at constructing an inflationary model in $N=1$ supergravity
proposed a generic form for the superpotential for a single inflaton \cite{nost},
the simplest form being $W = m^2 (1-a \Phi)^2$ \cite{hrr}.
As discussed in \cite{lw},  while this relatively simple model is capable of sufficient
inflation, it is an example of accidental inflation in the sense that the
coefficient of the linear term in the superpotential, $a$, must be extremely close to unity.
This model has also become one of Planck's casualties.
The scalar-to-tensor ratio in this model is very small,
but the value of $n_s$ predicted in this model is $n_s \simeq 1-4/N = 0.933$ for $N=60$~\cite{But},
since the effective potential varies insufficiently slowly.

In a supergravity model with a generic K\"ahler potential for the chiral
supermultiplets there are quadratic $|\phi|^2$ terms, which cause variations in the effective
potential that destroy its suitability for inflation, an obstacle known as the $\eta$-problem~\cite{inflationreviews}.
As was pointed out in~\cite{EENOS}, a natural solution to this problem is
offered by no-scale supergravity~\cite{no-scale}, whose motivations were summarized earlier.
In such a model, quadratic terms are
suppressed, and the effective scalar potential resembles that in a globally-supersymmetric model,
thanks to an underlying non-compact SU(N,1)/SU(N) $\times$ U(1)
symmetry. 

Other no-scale supergravity approaches have also been proposed \cite{othernoscale}, as
well as models based on 
a non-compact Heisenberg symmetry~\cite{BG}, a shift symmetry~\cite{Y,Davis:2008fv,klor},
or string theory \cite{Silverstein:2008sg}. 
The SU(N,1) model \cite{EENOS} was based on the superpotential $W = m^2(\phi - \phi^4/4)$
and gives similar predictions for the inflationary parameters as the minimal $N=1$ model discussed above.
This too is an example of accidental inflation \cite{lw} and a small change in the coefficient of the quartic term
would lead to parameters consistent with Planck data \cite{Planckinflation}.

In this paper we show how one can elevate the simplest globally-supersymmetric
Wess-Zumino inflationary model of~\cite{Croon} to a no-scale supergravity
version (NSWZ). Concretely, we study a model in which the inflaton superfield is
embedded in an SU(2,1)/SU(2) $\times$ U(1) no-scale supergravity sector together with
a modulus field $T$ (which we assume to be fixed by other dynamics~\cite{ENO7})
and find a range of the parameters where it is compatible with
the Planck data~ \cite{Planckinflation}. Quite remarkably, as we show, the NSWZ model is
the conformal equivalent of an $R + R^2$ model of gravity for one specific value of ${\hat \mu}/\lambda$,
so that in this case our realization of inflation in the NSWZ model is equivalent to the Starobinsky model of inflation~\cite{R2}. Thus we embed this model in a broader and attractive theoretical framework.

We first recall the basic relevant formulae governing the kinetic term
and the effective potential of scalar fields $\phi$ in ${\cal N} = 1$ supergravity,
specializing to the no-scale case with non-compact SU(N, 1)/SU(N) $\times$ U(1) symmetry.
The scalar sector may be characterized in general by a hermitian K\"ahler 
function $K$ and a holomorphic superpotential $W$ via the combination
$G \equiv K + \ln W + \ln W^*$. The kinetic term is then given by 
$K_i^{j^*} \partial_\mu \phi^i \partial \phi^*_j$, where the K\"ahler metric
$K_i^{j^*} \equiv \partial^2 K / \partial \phi^i \partial \phi^*_{j}$, and the
effective potential is
\begin{equation}
V \; = \; e^G \left[ \frac{\partial G}{\partial  \phi^i} K^i_{j^*}  \frac{\partial G}{\partial  \phi^*_j} - 3 \right] \, ,
\label{effpot}
\end{equation}
where $K^i_{j^*}$ is the inverse of the K\"ahler metric $K_i^{j^*}$.

In the minimal no-scale SU(2, 1)/SU(2) $\times$ U(1) case, there are two complex
scalar fields: $T$, a modulus field, and $\phi$, which we identify as the
inflaton field, with the K\"ahler function $K = -3 \ln ( T + T^* - |\phi|^2/3)$.
In this case, the kinetic terms
for the scalar fields $T$ and $\phi$ become
\begin{eqnarray}
{\cal L}_{KE} \; & = & \; \left( \partial_\mu \phi^*, \partial_\mu T^* \right) \left(\frac{3}{(T + T^* - |\phi|^2/3)^2} \right) \nonumber \\
 & & \left( \begin{array}{cc} (T + T^*)/3 & - \phi/3 \\ - \phi^*/3 & 1 \end{array} \right)
\left( \begin{array} {c} \partial^\mu \phi \\ \partial^\mu T \end{array} \right) \, ,
\label{no-scaleL}
\end{eqnarray}
and the effective potential becomes
\begin{equation}
V \; = \; \frac{{\hat V}}{(T + T^* - |\phi|^2/3)^2} : {\hat V} \; \equiv \; \left| \frac{\partial W}{\partial \phi} \right|^2 \, .
\label{effV}
\end{equation}
In early no-scale models \cite{EENOS,BG} it was assumed that $K$ was
fixed so that the potential up to a re-scaling was simply $\hat{V}$. 
Here we assume that the $T$ field has a
vacuum expectation value (vev) $2 \langle Re T \rangle = c$ and $\langle Im T \rangle = 0$
that is
determined by non-perturbative high-scale dynamics~\cite{ENO7}, as in
the K\"ahler correction provided in \cite{ekn3}.
In this case, we may neglect the kinetic mixing between the
$T$ and $\phi$ fields in (\ref{no-scaleL}), and are left with
the following effective Lagrangian for the inflaton field $\phi$: \begin{equation}
{\cal L}_{eff} \; = \; \frac{c}{(c - |\phi|^2/3)^2} |\partial_\mu \phi |^2 - \frac{{\hat V}}{(c - |\phi|^2/3)^2} \, .
\label{phiL}
\end{equation}
We assume as in~\cite{Croon} the minimal 
Wess-Zumino superpotential (\ref{WZW}) for the inflaton field.

\begin{figure}[h!]
\vskip -2in
\includegraphics[scale=0.5]{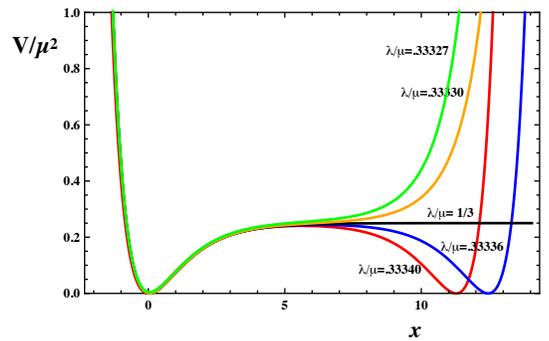}
\vskip -1.8in
\caption{\it The potential $V$ in the NSWZ model for choices of $\lambda \sim \mu/3$ 
in Planck units, as indicated.}
\label{pot}
\end{figure}

To better study the potential for the inflaton, we first transform $\phi$ to the field $\chi$:
\beq
\phi = \sqrt{3c} \tanh \left( \frac{\chi}{\sqrt{3}} \right) \, .
\eeq
With this field redefinition, the Lagrangian
becomes
\begin{eqnarray}
&&{\cal L}_{eff}  = {\rm sech}^2((\chi-\chi^*)/\sqrt{3}) \left[ | \partial_\mu \chi|^2  \right. - \\
& & \, \, \left. (\frac{3}{c}) \left|\sinh(\chi/\sqrt{3}) \left( \hat \mu \cosh(\chi/\sqrt{3})-\sqrt{3c}\lambda \sinh(\chi/\sqrt{3})\right) \right|^2 \right]  \, . \nonumber
\end{eqnarray}
Clearly the vev of the $T$ field can be absorbed into the definition of the mass
and, writing $\hat \mu = \mu\sqrt{c/3}$, the potential becomes
\beq
V = \mu^2\left |\sinh(\chi/\sqrt{3}) \left( \cosh(\chi/\sqrt{3})-\frac{3\lambda}{\mu} \sinh(\chi/\sqrt{3}) \right) \right|^2 \, .
\eeq
Writing $\chi$ in terms of its real and imaginary parts:
$\chi = (x + iy)/\sqrt{2}$ and, for reasons which will become clear,
considering the specific case where the quartic coupling $\lambda = \mu/3$ (in Planck units),
we have
\begin{eqnarray}
&& {\cal L}_{eff}  =  \frac{1}{2}\sec^2(\sqrt{2/3} y) \left( (\partial_\mu x)^2 + (\partial_\mu y)^2 \right) - \\
& & \,\, \mu^2 \frac{e^{-\sqrt{2/3}x}}{2} \sec^2(\sqrt{2/3}y)\left(\cosh{\sqrt{2/3}x})-\cos{\sqrt{2/3}y}\right) \, . \nonumber
\end{eqnarray}
The imaginary part of the inflaton is fixed to $y=0$ by the potential, having a mass 
$m_y = \mu/\sqrt{3}$ during inflation when $x$ is large and $m_y = \mu/\sqrt{6}$ 
at the end of inflation when $x=0$. Thus we expand the Lagrangian about 
$y=0$, in which case we have minimal kinetic terms for $x$ and $y$, accompanied by derivative
interaction terms. The potential for the real part of the inflaton now takes the form 
\beq
V = \mu^2 e^{-\sqrt{2/3}x} \sinh^2(x/\sqrt{6}) \, .
\label{nswzpot}
\eeq
This potential is depicted in Fig.~\ref{pot}, where we also display the potential
for values of $\lambda$ slightly perturbed from the nominal value of $\mu/3$.

We use the standard slow-roll expressions for the
tensor-to-scalar ratio $r$ and the spectral index $n_s$ for the scalar perturbations
in terms of the slow-roll inflation parameters $\epsilon, \eta$~\cite{inflationreviews}, which we evaluate in terms
of the canonically-normalized field $x$. 
In the NSWZ model described above the vev of $T$ is absorbed in the definition of the mass parameter
$\mu$, which is determined by the normalization of the quadrupole.
For the special case $\lambda = \mu/3$, we have
\beq
A_s = \frac{V}{24 \pi^2 \epsilon} =  \frac{ \mu^2}{8\pi^2} \sinh^4 (x/\sqrt{6}) \, ,
\eeq
implying a value $\mu = 2.2 \times 10^{-5}$ in Planck units for $N = 55$:
$\mu$ varies between 1.8 - 3.4 $\times 10^{-5}$ over the range of NSWZ models considered here.
Setting the remaining NSWZ parameter $\lambda = \mu/3$, we have
\begin{eqnarray}
\epsilon & = & \frac{1}{3} {\rm csch}^2(x/\sqrt{6})e^{-\sqrt{2/3} x} \, , \\
\eta & = & \frac{1}{3} {\rm csch}^2(x/\sqrt{6}) \left( 2e^{-\sqrt{2/3} x} -1 \right) \, ,
\end{eqnarray}
which allows us to determine the quantities $(n_s, r)$, once the value of the field $x$ is
fixed by requiring $N=50-60$ e-folds. The nominal choice of $N=55$ yields
$x=5.35$, $n_s = 0.965$, and $r = .0035$. 

\begin{figure}[h!]
\vskip -1.8in
\begin{center}
\includegraphics[scale=0.5]{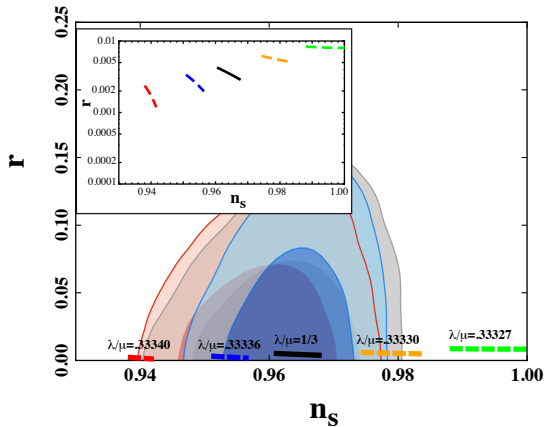}
\end{center}
\vskip -1.8in
\caption{\it Predictions from the NSWZ model
for the tilt $n_s$ in the spectral index of scalar perturbations and for the
tensor-to-scalar ratio $r$, compared with the 68 and 95\%
CL regions found in analyses of Planck and other data~\cite{Planckinflation}. 
In the main panel the lines are labelled by the values of $\lambda/\mu$ (in Planck units) assumed
in each case. In the inset, the same cases are shown on a log scale to display better the values of $r$.}
\label{fig:sugrainf}
\end{figure}

Fig.~\ref{fig:sugrainf} displays the predictions for $(n_s, r)$ of the NSWZ model
for five choices of the coupling $\lambda$ that yield $n_s \in [0.93,1.00]$ and $N \in [50, 60]$. 
The last 50-60 e-folds of inflation arise as $x$ rolls to zero from $\sim 5.1 - 5.8$,
the exact value depending on $\lambda$ and $N$.
As one can see, the values of $\lambda$ are constrained to be close to the critical value $\mu/3$,
for which we find
extremely good agreement with the Planck determination of $n_s$.  
The values of $r$ are rather small for $\lambda = \mu/3$, varying over the range 0.0012 -- 0.0084,
in the models considered.

At first sight, this success might appear to be another example of accidental inflation~\cite{lw} but, as we now show, this choice of $\lambda$ has
a more profound geometric interpretation.  The alert reader may have noticed resemblances of both the potential shown in Fig.~\ref{pot} and
the values of $(n_s,r)$ found for the $\lambda = \mu/3$ model with results for inflation in the $R+R^2$ model proposed by Starobinsky~\cite{R2}.
To probe further this resemblance, we examine the generalization of the Einstein-Hilbert action to contain an $R^2$ contribution,
where $R$ is the scalar curvature,
\beq
S=\frac{1}{2} \int d^4x \sqrt{-g} (R+R^2/6M^2) \, ,
\eeq
where $M \ll M_P$ is some mass scale.
This theory is conformally equivalent to canonical gravity plus a scalar field $\varphi$ \cite{whitt}.
Making the transformation $\tilde{g}_{\mu\nu} = (1 + \varphi/3M^2) g_{\mu\nu}$
and the field redefinition $\varphi^\prime = \sqrt{\frac{3}{2}} \ln \left( 1+ \frac{\varphi}{3 M^2} \right)$,
we obtain the action 
\beq
S=\frac{1}{2} \int d^4x \sqrt{-\tilde{g}} \left[\tilde{R} + (\partial_\mu \varphi^\prime)^2 - \frac{3}{2} M^2 (1- e^{-\sqrt{2/3}\varphi^\prime})^2 \right] \, ,
\eeq
corresponding to a potential 
\beq 
V =  \frac{3}{4} M^2 (1- e^{-\sqrt{2/3}\varphi^\prime})^2 \, .
\label{r2pot}
\eeq
The potential (\ref{r2pot}) is identical with the potential (\ref{nswzpot})
along the real direction of the NSWZ model!~\cite{Cecotti}
Moreover, we have the identification $M^2 = \mu^2/3$, which $= \hat \mu^2$ for $c =\langle (T + T^*) \rangle = 1$.
Thus the Starobinsky mass $M$ is directly related to the NSWZ mass $\hat \mu$ in the superpotential (\ref{WZW}).
We note that similar potentials are also obtained in Higgs Inflation and related models \cite{HI}.

We have shown in this paper that the simplest SU(2,1)/SU(2) $\times$ U(1) no-scale supergravity
model with a single modulus field $T$ and a single matter field $\phi$ with
the simplest renormalizable Wess-Zumino superpotential, identified with
the inflaton, is capable of yielding cosmological inflation with values of the scalar spectral
tilt $n_s$ and the tensor-to-scalar ratio $r$ within the region favoured by Planck and other data at the 68\% CL.
Successful inflation is obtained for $\lambda \simeq \mu/3$ in Planck units.
This NSWZ model is a proof of the existence of acceptable models of
inflation based on no-scale supergravity, and normally we would not advocate
that its details should necessarily be taken literally. For example, a
realistic no-scale model derived from a generic compactification of
string theory would have more moduli fields, with many matter
fields that could be the inflaton, with a superpotential more complicated than assumed
here.

However, it is truly striking that the NSWZ model is conformally equivalent to the Starobinsky $R^2$ model~\cite{R2}
for the specific choice $\lambda = \mu/3$ in Planck units. This correspondence suggests that there is
a profound geometric interpretation of this model that remains to be understood.

\noindent {\bf Acknowledgements. }  J.E. thanks Djuna Croon, Mirjam Cveti{\v c}, 
Nick Mavromatos, Henry Tye and Gary Shiu for discussions, and K.A.O. thanks
Andrei Linde and Misha Voloshin for discussions.
The work of J.E. was supported in part by
the London Centre for Terauniverse Studies (LCTS), using funding from
the European Research Council 
via the Advanced Investigator Grant 267352.
The work of D.V.N. was supported in part by the
DOE grant DE-FG03-95-Er-40917.
The work of K.A.O. was supported in part
by DOE grant DE--FG02--94ER--40823 at the University of Minnesota.


\vspace{1cm}


\begin{thebibliography}{99}






\bibitem{Planckinflation}
P.~A.~R.~Ade {\it et al.}  [Planck Collaboration],
  arXiv:1303.5082 [astro-ph.CO].
  
  \bibitem{oliverev}
  K.~A.~Olive,
  Phys.\ Rept.\  {\bf 190} (1990) 307.

  
\bibitem{inflationreviews}
See, for example:
A. D. Linde, {\it Particle  
Physics and
Inflationary Cosmology} (Harwood, Chur, Switzerland, 1990); 
  D.~H.~Lyth and A.~Riotto,
{\it Phys.\ Rep.}  {\bf 314} (1999) 1
[arXiv:hep-ph/9807278].
  J.~Martin, C.~Ringeval and V.~Vennin,
  arXiv:1303.3787 [astro-ph.CO].
  
  
  
\bibitem{R2}
A.~A.~Starobinsky,
  Phys.\ Lett.\ B {\bf 91}, 99 (1980);
   V.~F.~Mukhanov and G.~V.~Chibisov,
  JETP Lett.\  {\bf 33}, 532 (1981)
  [Pisma Zh.\ Eksp.\ Teor.\ Fiz.\  {\bf 33}, 549 (1981)];
  A.~A.~Starobinsky,
  Sov.\ Astron.\ Lett.\  {\bf 9}, 302 (1983).
  
  \bibitem{HI}
  F.~Bezrukov and M.~Shaposhnikov,
 JHEP {\bf 0907}, 089 (2009)
 [arXiv:0904.1537 [hep-ph]].
 
 
 
 \bibitem{others}
A.~Linde, M.~Noorbala and A.~Westphal,
 JCAP {\bf 1103}, 013 (2011)
 [arXiv:1101.2652 [hep-th]];
  S.~Ferrara, R.~Kallosh, A.~Linde, A.~Marrani and A.~Van Proeyen,
 Phys.\ Rev.\ D {\bf 83} (2011) 025008
 [arXiv:1008.2942 [hep-th]].
 
 \bibitem{ENOT}
J.~R.~Ellis, D.~V.~Nanopoulos, K.~A.~Olive and K.~Tamvakis,
  Phys.\ Lett.\ B {\bf 118} (1982) 335;
  Phys.\ Lett.\ B {\bf 120} (1983) 331;
  Nucl.\ Phys.\ B {\bf 221} (1983) 524.

\bibitem{sugra}
D.~Z.~Freedman, P.~van Nieuwenhuizen and S.~Ferrara,
  Phys.\ Rev.\ D {\bf 13} (1976) 3214;
  S.~Deser and B.~Zumino,
  Phys.\ Lett.\ B {\bf 62} (1976) 335.

\bibitem{no-scale}
E.~Cremmer, S.~Ferrara, C.~Kounnas and D.~V.~Nanopoulos,
  Phys.\ Lett.\ B {\bf 133} (1983) 61;
  J.~R.~Ellis, C.~Kounnas and D.~V.~Nanopoulos,
  Nucl.\ Phys.\ B {\bf 247} (1984) 373;
  A.~B.~Lahanas and D.~V.~Nanopoulos,
  Phys.\ Rept.\  {\bf 145} (1987) 1.

  
  \bibitem{Witten}
E.~Witten,
  Phys.\ Lett.\ B {\bf 155} (1985) 151.
  
  \bibitem{MC}
   O.~Buchmueller, R.~Cavanaugh, M.~Citron, A.~De Roeck, M.~J.~Dolan, J.~R.~Ellis, H.~Flacher and S.~Heinemeyer {\it et al.},
  Eur.\ Phys.\ J.\ C {\bf 72}, 2243 (2012)
  [arXiv:1207.7315].

  \bibitem{KLtest}
  R.~Kallosh and A.~D.~Linde,
 JCAP {\bf 0704} (2007) 017
 [arXiv:0704.0647 [hep-th]].
  
\bibitem{Croon}
D.~Croon, J.~Ellis and N.~E.~Mavromatos,
  arXiv:1303.6253 [astro-ph.CO].
  
  \bibitem{ab}
  A.~D.~Linde,
 Phys.\ Lett.\ B {\bf 132}, 317 (1983);
 A.~D.~Linde,
Pisma Zh.Eksp.Teor.Fiz. 37, 606  (1983) [JETP Lett. 37, 724 (1983)].
   A.~Albrecht and R.~H.~Brandenberger,
  Phys.\ Rev.\ D {\bf 31}, 1225 (1985).
   
  
  
  \bibitem{nos}
D.~V.~Nanopoulos, K.~A.~Olive and M.~Srednicki,
  Phys.\ Lett.\  B {\bf 127}, 30 (1983).

    
\bibitem{WZ}
J.~Wess and B.~Zumino,
  Nucl.\ Phys.\ B {\bf 70} (1974) 39.
  
  
  \bibitem{nost}
   D.~V.~Nanopoulos, K.~A.~Olive, M.~Srednicki and K.~Tamvakis,
  Phys.\ Lett.\ B {\bf 123}, 41 (1983).
  
  \bibitem{hrr}
   R.~Holman, P.~Ramond and G.~G.~Ross,
  Phys.\ Lett.\ B {\bf 137}, 343 (1984).
  
  \bibitem{lw}
   A.~D.~Linde and A.~Westphal,
  JCAP {\bf 0803}, 005 (2008)
  [arXiv:0712.1610 [hep-th]].
  
\bibitem{But}
 This model still falls within the region allowed by Planck at the 95\% CL 
for $N=70$ or a slight deviation in $a$ from unity, by 1 part in $10^6$.
    
\bibitem{EENOS}
J.~R.~Ellis, K.~Enqvist, D.~V.~Nanopoulos, K.~A.~Olive and M.~Srednicki,
  Phys.\ Lett.\ B {\bf 152} (1985) 175
   [Erratum-ibid.\  {\bf 156B} (1985) 452].
  
  
  
  \bibitem{othernoscale}
    A.~S.~Goncharov and A.~D.~Linde,
  Class.\ Quant.\ Grav.\  {\bf 1},  L75 (1984).


\bibitem{BG}
P.~Binetruy and M.~K.~Gaillard,
  Phys.\ Lett.\ B {\bf 195} (1987) 382;
  H.~Murayama, H.~Suzuki, T.~Yanagida and J.~Yokoyama,
  Phys.\ Rev.\  D {\bf 50}, 2356 (1994)
  [arXiv:hep-ph/9311326];
  S.~Antusch, M.~Bastero-Gil, K.~Dutta, S.~F.~King and P.~M.~Kostka,
  Phys.\ Lett.\ B {\bf 679} (2009) 428
  [arXiv:0905.0905 [hep-th]].
  
\bibitem{Y}
M.~Kawasaki, M.~Yamaguchi and T.~Yanagida,
  Phys.\ Rev.\ Lett.\  {\bf 85} (2000) 3572
  [hep-ph/0004243];
  K.~Nakayama, F.~Takahashi and T.~T.~Yanagida,
  arXiv:1303.7315 [hep-ph].
  
  
\bibitem{Davis:2008fv}
  S.~C.~Davis and M.~Postma,
  JCAP {\bf 0803}, 015 (2008)
  [arXiv:0801.4696 [hep-ph]].
  
  \bibitem{klor}
   R.~Kallosh and A.~Linde,
  JCAP {\bf 1011}, 011 (2010)
  [arXiv:1008.3375 [hep-th]];
    R.~Kallosh, A.~Linde and T.~Rube,
  Phys.\ Rev.\  D {\bf 83}, 043507 (2011)
  [arXiv:1011.5945 [hep-th]];
 R.~Kallosh, A.~Linde, K.~A.~Olive and T.~Rube,
  Phys.\ Rev.\ D {\bf 84}, 083519 (2011)
  [arXiv:1106.6025 [hep-th]].
  
\bibitem{Silverstein:2008sg}
  E.~Silverstein and A.~Westphal,
  Phys.\ Rev.\  D {\bf 78}, 106003 (2008)
  [arXiv:0803.3085 [hep-th]];
  L.~McAllister, E.~Silverstein and A.~Westphal,
  Phys.\ Rev.\  D {\bf 82}, 046003 (2010)
  [arXiv:0808.0706 [hep-th]].
   X.~Dong, B.~Horn, E.~Silverstein and A.~Westphal,
  arXiv:1011.4521 [hep-th].
  

  


  
  
\bibitem{ENO7}
For examples showing that this is possible while preserving the
no-scale structure and reducing to our model in a suitable limit,
and hence justifying our subsequent analysis, see
J.~Ellis, D.~V.~Nanopoulos and K.~A.~Olive,
  arXiv:1307.3537 [hep-th].

  \bibitem{ekn3}
J.~R.~Ellis, C.~Kounnas and D.~V.~Nanopoulos,
  Phys.\ Lett.\ B {\bf 143}, 410 (1984).
  
  
\bibitem{whitt}
B.~Whitt,
  Phys.\ Lett.\ B {\bf 145}, 176 (1984).
  
\bibitem{Cecotti}
We thank A.~Kehagias for pointing out to us a previous derivation
of the $R + R^2$ model from a different version of no-scale SU(2, 1)/SU(2) $\times$ U(1) supergravity:
S.~Cecotti,
  Phys.\ Lett.\ B {\bf 190} (1987) 86.


\end{thebibliography}
\end{document}